\documentclass[12pt]{article}
\usepackage{amsfonts}
\usepackage[fleqn]{amsmath}
\usepackage{amsmath}
\usepackage{amssymb} 
\usepackage{hhline}
\usepackage{nicefrac}
\usepackage[toc]{appendix}
\usepackage{upgreek}
\usepackage{mathrsfs}
\usepackage[scale={.75,.75}]{geometry}
\usepackage[T1]{fontenc}
\usepackage{bm}
\usepackage{pstool}
\usepackage{epsf}
\usepackage{psfrag}
\usepackage{fleqn} 
\usepackage{mathrsfs} 
\usepackage{cite} 
\usepackage{longtable}
\usepackage{caption}
\usepackage{cite}
\usepackage{graphicx}
\usepackage{verbatim}

\numberwithin{equation}{section}
\numberwithin{table}{section}
\allowdisplaybreaks

\DeclareMathOperator{\Tr}{Tr}

\newcommand{\be}{\begin{equation}}
\newcommand{\bea}{\begin{array}}
\newcommand{\eea}{\end{array}}

\newcommand{\bes}{\begin{split}}
\newcommand{\ees}{\end{split}}

\newcommand{\tm}{y}
\newcommand{\e}{\epsilon}

\newcommand{\Boxq}{(\partial_{t_1}^2+\omega_{{\bf q}}^2)}
\newcommand{\Boxk}{(\partial_{t_1}^2+\omega_{{\bf k}}^2)}

\newcommand{\beq}{\begin{eqnarray}}
\newcommand{\eeq}{\end{eqnarray}}

\newcommand{\C}{\mathcal{C}}

\numberwithin{equation}{section}
\numberwithin{table}{section}
\allowdisplaybreaks

\begin{document}
\date{\mbox{\today }}

\title{ 
{\normalsize     
\hfill\mbox{}\\
\hfill\mbox{}\\}
\vspace{1cm}
\bf 
Backreaction Effects on Nonequilibrium Spectral Function \\[8mm]}
%
\author{{\bf\normalsize Sebasti\'an~Mendizabal$\dagger$ and Juan Cristobal Rojas*}\vspace{1cm}\\ 
{\it\normalsize $\dagger$ Universidad T\'ecnica Federico Santa Mar\'ia, Santiago, Chile}\\
{\it\normalsize *Universidad Cat\'olica del Norte, Antofagasta, Chile}
{\it\normalsize }
}

\maketitle

\thispagestyle{empty}


\begin{abstract}
\noindent

We show how to compute  the spectral function for a scalar theory in two different
scenarios: one which disregards back-reaction i.e. the response of the environment to 
the external particle, and the other one where back-reaction is considered.
The calculation was performed using the
Kadanoff-Baym equation through the Keldysh formalism. When back-reaction is
neglected, the spectral function is equal to the equilibrium one, which can be represented as a Breit-Wigner distribution. When back-reaction is introduced we observed a damping in the spectral function
of the thermal bath.  Such behavior modifies the damping rate for particles created
within the bath. This certainly implies phenomenological consequences right after the Big-Bang, when the
primordial bath was created.

\end{abstract}

\newpage
\section{Introduction}

The understanding of nonequilibrium phenomena has shown itself to be a very elusive issue over the years, yet it is crucial to comprehend these type of phenomena as it has mold our universe as it is. Several mechanisms has been introduced to deal with these issues, for example in the context of inflaton decay, electro-weak baryogenesis, leptogenesis, dark matter production, and Big-Bang nucleosynthesis among others (for reviews about these issues see \cite{kolb, weinberg}).

Nonequilibrium phenomena are an initial condition problem that can be treated with the evolution of the density matrix or the evolution of fields. As a matter of simplicity it is usually treated as a weakly coupled particle produced in a strongly coupled thermal bath. This is done to guarantee a power expansion in terms of the coupling constant between the interacting fields.

It has been shown that out-of-equilibrium processes there are always two independent important quantities, namely the spectral function and the statistical propagator, this is related to the fact that propagators will not only depend on the difference of time between two events but also on the ``center of mass'' time. In equilibrium, these two quantities are related via the Kubo-Martin-Schwinger relation \cite{KMS} , or the Fluctuation-Dissipation Theorem \cite{fluctuation,Das} . Conditions which does not apply in an out-of-equilibrium scenario.

The spectral function, as its names shows, have all the information of a (non-)thermal field. In equilibrium, and for small imaginary part of the self-energy, it can be described as a Breit-Wigner function. This is often called a quasi-particle, whose pole can be defined as the effective mass of a field in a thermal bath, and its width can be interpreted as the decay parameter which describes the production or annihilation. On the other hand, the statistical propagator is related to the occupation number of the fields, which normally tells us the number of the particles in the bath.

Commonly these two quantities are computed via the semi-classical Boltzmann equation with quantum corrections inserted in the collision factor. This type of approach lacks of several important physical aspects when one include phenomena as for example coherent oscillations, off-shell corrections or non-Markovian effects. In order to properly approach these issues, one needs to start with the well known Kadanoff-Baym equations (KBE) \cite{kadanoff}. The KBE are differential-integral equations where the important properties come from the convolution of full propagators with the self-energy. Although the KBE are hard to work with and normally they can only be treated numerically, it was shown in \cite{Anisimov} that they can be solved analytically in the leptogenesis scenario. This simple scenario can be used as the  starting point to comprehend the properties of the KBE and their relation to the semi-classical Boltzmann equation, where already numerous studies has been performed \cite{Buchmuller32, giudice, desimone2, garny1, garny2, garny3, koksma, drewes, Drewes:2010pf, Buchmuller:2011mw, Fidler:2011yq, Garbrecht:2011xw, Blanchet:2011xq, Garny:2011hg, Millington:2013ema, Gautier:2012vh, Frossard:2012pc, Millington:2013isa}. For example, it was shown in \cite{Drewes:2012qw} that the main difference between both evolution equations comes from the fact that there is a time where quantum processes are important, time that is related to the decay width of the out-of-equilibrium field.

In certain nonequilibrium scenarios backreaction can not be easily disregarded since it determines particle emission and absorption within the bath. The study of these effects can be approached in a methodic way using the results obtained in \cite{Anisimov} for a weakly coupled field. Results which can be extended for more complicated scenarios. In this context, backreaction can be treated as a 2-loop process, which means that quantum effects can have a big impact in the spectral function.

In this paper we will focus on a simple scalar model following the definitions of \cite{scalarmendi} with an interaction of the form $\mathcal{L}\sim g\phi\chi^2$, where the field $\phi$ is the weakly coupled out-of-equilibrium field and $\chi$ is a strongly coupled field that composes the thermal bath, which can also be coupled to other fields in equilibrium. As in \cite{Anisimov} and \cite{scalarmendi}, the mass of the $\phi$ field is much larger than the mass of the particles in the bath, i.e. $m_{\phi}\gg m_{\chi}$. Although, it can be generalized easily for different hierarchy of masses. After performing a suitable expansion of the KBE in the coupling constant $g$, one can analytically find the proper propagators of the fields and finally study the introduction of backreaction in a consistent way.

This paper will be constructed as follows, in chapter 2 we will give a brief review on equilibrium and nonequilibrium formalism for the calculation of the spectral function and the statistical propagator. In chapter 3 we follow the prescription and show the solution of these quantities when backreaction is neglected. In chapter 4 we include backreaction where we calculated the time evolution of the spectral function as it goes out of equilibrium and thermalizes later. In chapter 6 we discuss the new ``sum rule'' for the spectral function. In chapter 7 we give numerical solutions for the nonequilibrium spectral function.

\section{Keldysh Formalism for scalar fields}

To understand nonequilibrium phenomena we will first give a brief description of the Keldysh formalism already presented in several articles \cite{keldysh, leb96, ber04, csx84, zin93, yok04, bdh04}. A field $\Phi$ is coupled to a thermal bath 
described by a self-energy $\Pi_C$. The Green's function $\Delta_C$ satisfies the Schwinger-Dyson equation
\begin{equation}\label{sde}
(\square_1 +m^2)\Delta_C(x_{1},x_{2})+i\int_{C}d^{4}x' \Pi_{C}(x_{1},x')
\Delta_{C}(x',x_{2})=-i\delta_{C}(x_{1}-x_{2})\ ,
\end{equation}
where $\Box_1=(\partial^2/\partial x_1^2)$ and $m$ is the mass of the field $\Phi(x)$. The scalar propagator $\Delta_C$ and self-energy $\Pi_C$ are defined in the $x^0$-contour (Fig. \ref{contour1}) as
\begin{equation}\label{causal}
\Delta_C(x_1,x_2)
=\theta_C(x^0_1,x^0_2)\Delta^>(x_1,x_2) 
+ \theta_C(x^0_2,x^0_1)\Delta^<(x_1,x_2)\ ,
\end{equation}
\begin{equation}\label{causal}
\Pi_C(x_1,x_2)
=\theta_C(x^0_1,x^0_2)\Pi^>(x_1,x_2) 
+ \theta_C(x^0_2,x^0_1)\Pi^<(x_1,x_2)\ .
\end{equation}
The $\theta$-functions enforce path ordering along the contour $C$, and 
$\Delta^>$ and $\Delta^<$ are the \; correlation functions
\begin{eqnarray}
\Delta^>(x_1,x_2) &=& \langle \Phi(x_1)\Phi(x_2)\rangle =
\Tr(\hat{\rho}\Phi(x_1)\Phi(x_2))\ ,\label{forw}\\ 
\Delta^<(x_1,x_2) &=& \langle \Phi(x_2)\Phi(x_1)\rangle =
\Tr(\hat{\rho}\Phi(x_2)\Phi(x_1))\label{back}\ ,
\end{eqnarray}
where $\hat{\rho}=\exp[\beta(\mathcal{F}-\mathcal{\hat{H}})]$ is the density matrix of the system at some initial time $t_i$.  Here, $\beta$ is the inverse of the temperature,  $\mathcal{F}$ is the free
energy and $\mathcal{\hat{H}}$ the Hamiltonian. The self energies $\Pi^{\gtrless}$ are calculated appropriately.

\begin{figure}[t]\label{contour1}
  \centering
    \includegraphics[width=12cm]{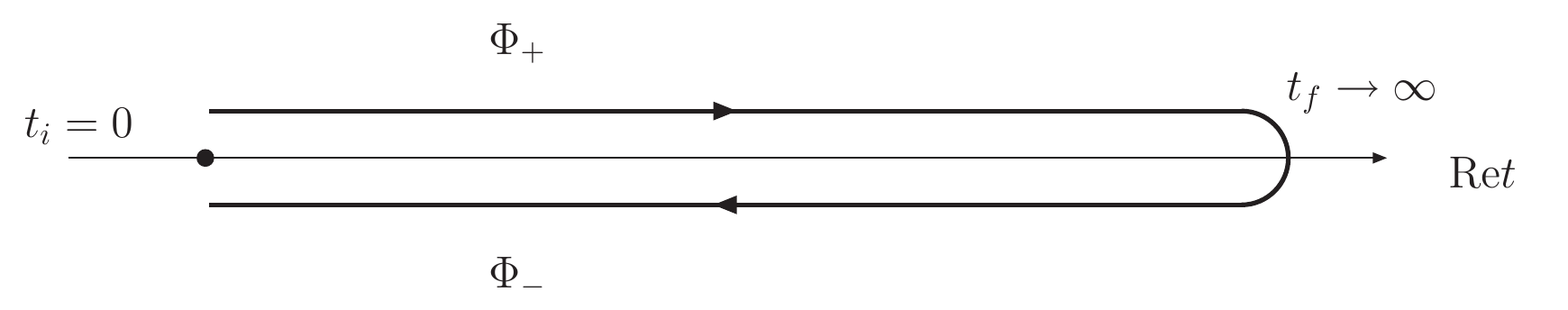}\label{contour1}
    \caption{Path in the complex time plane for nonequilibrium Green's 
functions.}\label{contour1}
\end{figure}
Introducing retarded and advanced representation for the propagators and self-energies
\begin{eqnarray}
\hat{O}^{R}(x_{1},x_{2})
=&&\theta(t_{1}-t_{2})(\hat{O}^{>}(x_{1},x_{2})
-\hat{O}^{<}(x_{1},x_{2}))\ ,
 \label{retarded}\\ 
\hat{O}^{A}(x_{1},x_{2})
=&-&\theta(t_{2}-t_{1})(\hat{O}^{>}(x_{1},x_{2})
-\hat{O}^{<}(x_{1},x_{2}))\ ,
\label{advanced}
\end{eqnarray}
\noindent where the symbol $\hat{O}$ denotes $\Delta$ or $\Pi$. From Eqs.~(\ref{retarded})-(\ref{advanced}) one obtains the
Kadanoff-Baym equations for the correlation functions $\Delta^>$ and $\Delta^<$,
\begin{align}
(\square_{1}+m^{2})\Delta^{>}(x_{1},x_{2})
&=-i\int d^{4}x' \left(\Pi^{>}(x_{1},x')\Delta^{A}(x',x_{2})
+ \Pi^{R}(x_{1},x')\Delta^{>}(x',x_{2})\right)\ ,\label{kadanoffbaym1}\\
(\square_{1}+m^{2})\Delta^{<}(x_{1},x_{2})
&=-i\int d^{4}x' \left(\Pi^{<}(x_{1},x')\Delta^{A}(x',x_{2})
+\Pi^{R}(x_{1},x')\Delta^{<}(x',x_{2})\right)\ .\label{kadanoffbaym2}
\end{align}
\noindent We can also define the real symmetric and antisymmetric form for correlation functions
and self-energies:
\begin{align}
\hat{O}^{+}(x_{1},x_{2})
&=\frac{1}{2}\big(\hat{O}^{>}(x_{1},x_{2})+\hat{O}^{<}(x_{1},x_{2})\big)\ ,
\label{pplus}\\
\hat{O}^{-}(x_{1},x_{2})
&=i\big(\hat{O}^{>}(x_{1},x_{2})-\hat{O}^{<}(x_{1},x_{2})\big)\ ,
\label{pminus}
\end{align}

\noindent which will determine the retarded and advanced operators,

\begin{equation}
\hat{O}^{R}(x_{1},x_{2})=\theta(t_{1}-t_{2})\hat{O}^{-}(x_{1},x_{2})\ ,\quad
\hat{O}^{A}(x_{1},x_{2})=-\theta(t_{2}-t_{1})\hat{O}^{-}(x_{1},x_{2})\ .
\label{retadvance}
\end{equation}

In the following we shall restrict ourselves to a system with spatial 
translational 
invariance. In this case all two-point functions only depend 
on the difference of spatial coordinates and it is 
convenient to perform a spatial Fourier transformation in ${\bf x}_1-{\bf x}_2$. Introducing definitions (\ref{pplus})-(\ref{retadvance}) into  (\ref{kadanoffbaym1}) and
(\ref{kadanoffbaym2}), one obtains
an homogeneous equation for $\Delta^-$ and an inhomogeneous equation for
$\Delta^+$
\begin{align}
\label{kbe1}
&\Boxk\Delta^{-}_{\bf{k}}(t_{1},t_{2})+  
\int_{t_{2}}^{t_{1}} dt'\Pi^{-}_{\bf{k}}(t_{1},t')\Delta^{-}_{\bf{k}}(t',t_{2})=0\ , \\
&\Boxk\Delta^{+}_{\bf{k}}(t_{1},t_{2}) 
+\int_{t_{i}}^{t_{1}} dt'\Pi^{-}_{\bf{k}}(t_{1},t')\Delta^{+}_{\bf{k}}(t',t_{2})
=\int_{t_{i}}^{t_{2}} dt' \Pi^{+}_{\bf{k}}(t_{1},t')\Delta^{-}_{\bf{k}}(t',t_{2})\ ,
\label{KB2}
\end{align}
\noindent where $\omega_{\bf{k}}^2={\bf{k}}^{2}+m^{2}$. 

We shall refer to these as equations as the first and second Kadanoff-Baym
equation respectively. $\Delta^{-}_{\bf k}$ and $\Delta^{+}_{\bf k}$ are known as spectral function and statistical 
propagator (cf.~\cite{ber04}). It is important to notice that although it appears that equation (\ref{kbe1}) does not depend on the initial time, the solution for $\Delta^{-}_{\bf k}$ will have a dependence on the initial conditions if back-reaction is considered in the self energy $\Pi^-_{\bf k}$.

Even though the starting system gives the initial conditions for the statistical propagator, the spectral function will always satisfy micro-causality

\begin{align}
\Delta^{-}(x_{1},x_{2})|_{t_{1}=t_{2}} &= 0\ , \label{con1}\\
\partial_{t_{1}}\Delta^{-}(x_{1},x_{2})|_{t_{1}=t_{2}}
=-\partial_{t_{2}}\Delta^{-}(x_{1},x_{2})|_{t_{1}=t_{2}}
 &= \delta({\bf x}_1-{\bf x}_2)\ , \label{con2}\\
\partial_{t_{1}}\partial_{t_{2}}
\Delta^{-}(x_{1},x_{2})|_{t_{1}=t_{2}} &= 0\ .
\label{con3}
\end{align}

\section{Solution of the spectral function and the statistical propagator without backreaction}
 
\begin{figure}[t]\label{feynman21}
  \centering
    \includegraphics[width=12cm]{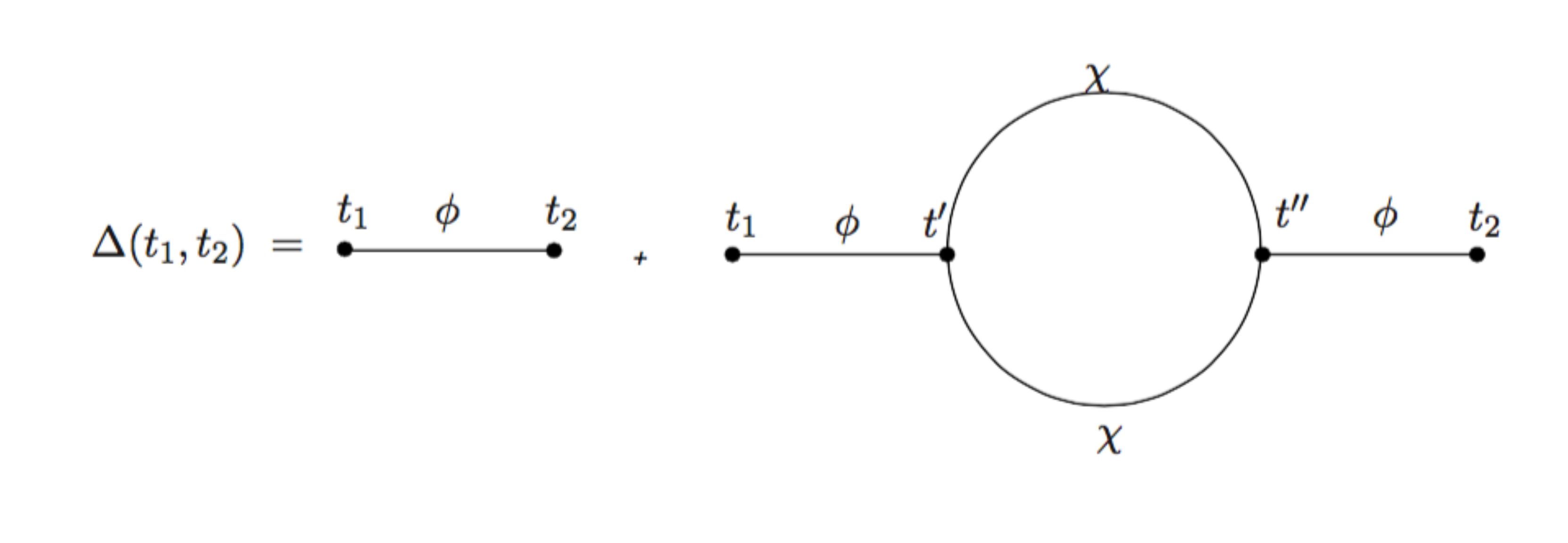}\label{feynman21}
    \caption{Diagrams contributing to the full propagator of $\phi$.}\label{feynman21}
\end{figure}

It was shown in \cite{scalarmendi} that the spectral function will only depend on the time difference $y=t_1-t_2$ when
backreaction is neglected. That is a direct consequence of time-translational invariance of the self-energy. Hence, the first Kadanoff-Baym equation (\ref{kbe1}) takes the form
\begin{equation}
\left(\partial_y^2 + \omega_{{\bf k}}^2\right)\Delta_{c}^-(y) + \int^{y}_{0} dy'
\Pi_{\bf k}^-(y-y')\Delta_{\bf k}^-(y') = 0\ .
\label{eqspectral}
\end{equation}

\noindent Its graphical representation is depicted in fig. \ref{feynman21}. The above integro-differential homogeneous equation can be easily solved by performing a Laplace transformation, $\tilde\Delta_{{\bf k}}^-(s)=\mathscr{L}\{\Delta^-_{{\bf k}}(y)\}$.
Since the spectral function $\Delta^-_{{\bf k}}(y)$ satisfies the boundary conditions 
(\ref{con1})-(\ref{con3}), we obtain
\begin{equation}
\Delta^-_{{\bf k}}(y) = \int_{\C_B}{ds\over 2\pi i}
{e^{s\tm}\over s^2+\omega^2_{{\bf k}}+\tilde\Pi^{-}_{{\bf k}}(s)}\ .
\label{brom}
\end{equation}

The subscript $\C_B$ refers to the Bromwich contour which is used to invert the Laplace transformation by using the analytical properties of the self-energy $\tilde\Pi^{-}(s)$. On the real axis $\tilde\Pi^{-}(s)$ is 
real, while on the parts of the contour which are parallel to the imaginary 
axis one has (for detail see \cite{bdh04}). 
\begin{equation}
\tilde\Pi^{-}(i\omega\pm\e) = {\rm Re}\Pi^R_{{\bf k}}(\omega)\pm 
i{\rm Im}\Pi^R_{{\bf k}}(\omega)\ ,
\end{equation}

\noindent with
\begin{equation}
{\rm Im} \Pi^{R}_{\bf{k}}(\omega)=\frac{1}{2i}
\left(\Pi^{R}_{\bf{k}}(\omega+i\epsilon)
-\Pi^{R}_{\bf{k}}(\omega-i\epsilon)\right)\ .
\end{equation}

\noindent Hence, the expression (\ref{brom}) can be written in terms of a real function $\rho_{{\bf k}}(\omega)$
\begin{equation}
\Delta^-_{{\bf k}}(\tm)=i\int^{\infty}_{-\infty}\frac{d\omega}{2\pi} 
e^{-i\omega \tm}\rho_{{\bf k}}(\omega)\ ,
\label{dmin}
\end{equation}

\noindent where the spectral function $\rho_{{\bf k}}(\omega)$, after renormalisation, is given in terms of real and 
imaginary part of the renormalized self-energy $\Pi^{R,r}_{{\bf k}}(\omega)$,  

\begin{equation}\label{spectralfunctionReno}
\rho^{r}_{{\bf k}}(\omega) = 
\frac{-2 {\rm Im}\Pi^{R,r}_{{\bf k}}(\omega)+2\omega\epsilon}
{\left(\omega^{2}-\omega_{{\bf k}}^{2}-{\rm Re}\Pi_{{\bf k}}^{R,r}(\omega)\right)^2
+\left({\rm Im}\Pi^{R,r}_{{\bf k}}(\omega)+\omega\epsilon\right)^{2}}\ .
\end{equation}

The divergencies of spectral function and statistical propagator can be
removed in the same way by mass and wave function renormalisation at zero
temperature. In the following we shall drop the superscript `$r$' to keep the
notation simple.

The spectral function describes a quasi-particle resonance at finite 
temperature with energy $\Omega_{{\bf k}}$,
\begin{equation}
\Omega^2_{{\bf k}}-\omega_{{\bf k}}^{2}-{\rm Re}\Pi^{R}_{{\bf k}}(\Omega_{{\bf k}}) = 0\ ,
\end{equation}

\noindent and decay width defined with the imaginary part of the self-energy at $\omega=\Omega_{{\bf k}}$
\begin{equation}
\Gamma_{{\bf k}} \simeq -\frac{1}{\Omega_{{\bf k}}}\text{Im}\Pi^{R}_{{\bf k}}(\Omega_{{\bf k}})\ .
\end{equation}

\noindent With these new variables we get
\begin{equation}
\rho_{{\bf k}}(\omega) = 
\frac{2\Omega_{{\bf k}} \Gamma_{{\bf k}}}
{\left(\omega^{2}-\Omega_{{\bf k}}\right)^2
+\Omega_{{\bf k}}^2\Gamma_{{\bf k}}^2} {\rm sign}(\omega)\ ,
\end{equation}

\noindent and in $y$-space
\begin{equation}
\Delta^-_{{\bf k}}(\tm)=\frac{1}{\Omega_{{\bf k}}}\sin(\Omega_{{\bf k}}y)e^{-\frac{\Gamma_{{\bf k}}|y|}{2}}\ .
\end{equation}

\noindent  At zero temperature or big mass $M\gg {\rm Re}\Pi^{R}_{{\bf k}}(\Omega_{{\bf k}}) $
\begin{equation}
\Omega^2_{{\bf k}}|_{T=0} = \omega^2_{{\bf k}}\ .
\end{equation}
\noindent For simplicity we have neglected the effect of $\text{Im}\Pi^{R}_{{\bf k}}$ 
on the quasi-particle energy as it is much smaller than the real part of the self-energy ($\text{Im}\Pi^{R}_{{\bf k}} \ll \text{Re}\Pi^{R}_{{\bf k}}$).



We are now ready to solve the second Kadanoff-Baym equation (\ref{KB2}) for 
the statistical propagator, which for initial time $t_i=0$ is given by
\begin{equation}
\Boxk\Delta_{\bf k}^+(t_1,t_2)+\int^{t_1}_{0} dt'
\Pi_{\bf k}^-(t_1-t')\Delta_{\bf k}^+(t',t_2)=\zeta(t_1,t_2)\label{kb2}\ ,
\end{equation}

\noindent with
\begin{equation}
\zeta(t_1,t_2)=\int^{t_2}_{0} dt'
\Pi_{{\bf k}}^+(t_1-t')\Delta_{\bf k}^-(t'-t_2)\ .
\end{equation}
One easily verifies that the solution can be expressed as
\begin{equation}
\Delta^+_{{\bf k}}(t_1,t_2)=\hat{\Delta}_{{\bf k}}^+(t_1,t_2) 
+\Delta_{{\bf k},mem}^{+}(t_1,t_2)\ ,
\label{sol2}
\end{equation}
where $\hat{\Delta}_{{\bf k}}^+(t_1,t_2)$ satisfies the homogeneous equation
\begin{equation}\label{homoequa}
\Boxk\hat{\Delta}_{{\bf k}}^+(t_1,t_2)+\int^{t_1}_{0} dt'
\Pi_{{\bf k}}^-(t_1-t')\hat{\Delta}_{{\bf k}}^+(t',t_2) = 0\ ,
\end{equation}
and $\Delta_{{\bf k},mem}^{+}(t_1,t_2)$ satisfies the inhomogeneous memory integral and is given by
\begin{equation}
\Delta^{+}_{{\bf k},\text{mem}}(t_{1},t_{2}) =
\int_{0}^{t_{1}}dt'\int_{0}^{t_{2}}dt''
\Delta^{-}_{{\bf k}}(t_{1}-t')\Pi^{+}_{{\bf k}}(t'-t'')\Delta^{-}_{{\bf k}}(t''-t_{2})\ ,
\end{equation}
which can be written
\begin{eqnarray}
\Delta^{+}_{{\bf k},\text{mem}}(t_{1},t_{2}) =&&\int\frac{d\omega}{2\pi}
\left(\int_{0}^{t_{1}}dy_1\Delta^{-}_{{\bf k}}(y_1)e^{i\omega y_1}\right)\Pi^{+}_{{\bf k}}(\omega)\nonumber\\
&&\hspace{1cm}\times\left(\int_{0}^{t_{2}}dy_2\Delta^{-}_{{\bf k}}(-y_{2})e^{-i\omega y_2}\right)e^{-i\omega y}\ .
\end{eqnarray}
\noindent The self-energy satisfies the KMS condition
\begin{equation}
\Pi_{\bf
k}^{+}(\omega)=-\frac{i}{2}\coth\left(\frac{\beta\omega}{2}\right)
\Pi_{\bf k}^{-}(\omega)\ .
\end{equation}
Performing the time integral we get 
\begin{align}
\int_{0}^{t}dy e^{i\omega y}\Delta_{\bf k}^{-}(y)\ &=
\frac{1}{\omega_{\textbf{k}}^{2}-(\omega+i\Gamma_{\bf k}/2)^{2}} \\
&\times\frac{1}{\omega_{\textbf{k}}}
\big[i\omega\big(\sin(\omega_{\textbf{k}}t)
-\omega_{\textbf{k}}\cos(\omega_{\textbf{k}}t)\big)
e^{i(\omega+i\Gamma_{\bf k}/2)t}
+\omega_{\textbf{k}}\big]\ , \nonumber\\
\int_{0}^{t}dy e^{-i\omega y}\Delta_{\bf k}^{-}(-y)\ &=
\frac{1}{\omega_{\textbf{k}}^{2}-(\omega-i\Gamma_{\bf k}/2)^{2}} \\
&\times\frac{1}{\omega_{\textbf{k}}}
\big[i\omega\big(\sin(\omega_{\textbf{k}}t)
-\omega_{\textbf{k}}\cos(\omega_{\textbf{k}}t)\big)
e^{i(\omega+i\Gamma_{\bf k}/2)t}
+\omega_{\textbf{k}}\big]\ .\nonumber
\end{align}
One can now perform the $\omega$ integration using the residue theorem. Notice that the contour must be chosen according to the sign of the exponentials. Now, using the fact that 
\begin{equation}\label{gammadef}
\Gamma_{{\bf k}} \simeq -\frac{1}{2\Omega_{{\bf k}}}\Pi^{-}_{{\bf k}}(\Omega_{{\bf k}})\ ,
\end{equation} 
we finally get
\begin{align}
 \label{Gmem}
\Delta^{+}_{\textbf{k},{\rm mem}}(t,y) =
\frac{1}{2\omega_{\textbf{k}}}\coth\left(\frac{\beta\omega_{\textbf{k}}}{2}\right)
\cos(\omega_{\textbf{k}}y)\left(e^{-\frac{\Gamma_{{\bf k}}}{2}|y|}
-e^{-\Gamma_{{\bf k}} t}\right)\ .\nonumber
\end{align}
One can check that for $\Gamma_{\bf k}t\rightarrow\infty$ the KMS condition is satisfied. The full solution is given by the sum of the memory solution to the equilibrium one. The latter can be found with equation (\ref{homoequa}) with the vacuum initial conditions for the statistical propagator
\begin{equation}
\hat{\Delta}_{{\bf k}}^+(0,0)=\frac{1}{2\omega_{\bf k}}\ .
\end{equation}
Leading us to the final solution for the statistical propagator
\begin{align}
\Delta^{+}_{ \textbf{k}}(t,y)=\frac{1}{\omega_{ \textbf{k}}}\cos(\omega_{
\textbf{k}}y)\left[\frac{1}{2}\coth\left(\frac{\beta\omega_{ \textbf{k}}}{2}\right)e^{-\frac{\Gamma_{\bf k}}{2}|y|}-f^B(\omega_{ \textbf{k}})e^{-\Gamma_{ \textbf{k}}t}\right]\;.
\end{align}
Fot fixed $\Gamma_{ \textbf{k}}y$ and $\Gamma_{ \textbf{k}}t\rightarrow \infty$ one regains the expected equilibrium propagator.

\section{Nonequilibrium spectral function with backreaction}
\begin{figure}[t]\label{feynman1}
  \centering
    \includegraphics[width=15cm]{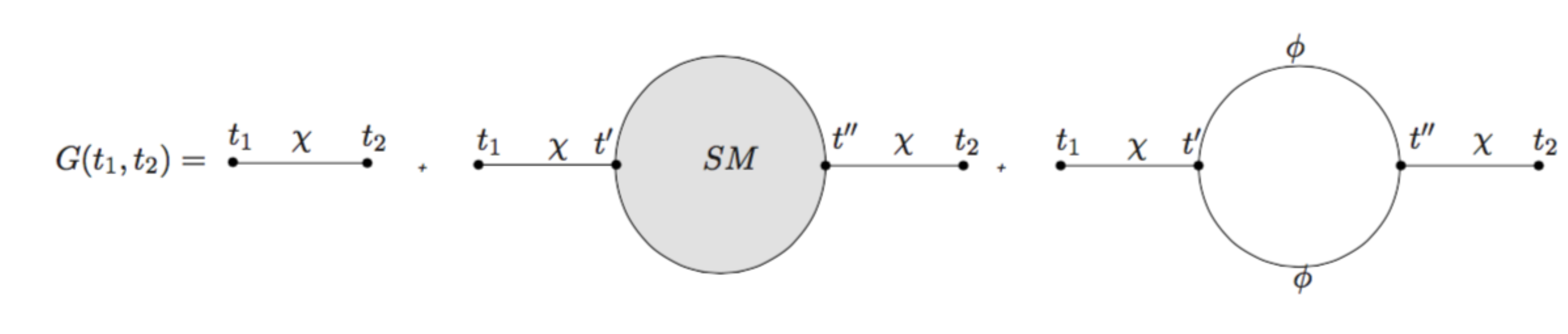}
    \caption{Self-energies diagrams with the incorporation of backreaction for the thermal bath. The first self-energy includes all possible equilibrium interaction. The last diagram include the nonequilibrium propagators of $\phi$.}\label{feynman1}
\end{figure}

To study backreaction one has to describe the effects that the bath has on itself. Taking into account that equilibrium propagators will only give equilibrium spectral functions, we need to include nonequilibrium propagators from the last section in the bath's self-energy $\Sigma_{\bf q}$. This is analogous to a 2-loop expansion in the Kadanoff-Baym equations. The spectral function of the bath $G^{-}_{\bf{q}}$ satisfies following equation
\begin{align}
\label{kbeG1}
\Boxq G^{-}_{\bf{q}}(t_{1},t_{2})+  
\int_{t_{2}}^{t_{1}} dt'\Sigma^{-}_{\bf{q}}(t_{1},t')G^{-}_{\bf{q}}(t',t_{2})=0\ .
\end{align}

For any interaction, one can write the spectral function as 

\begin{equation}
G_{{\bf q}}^-(t_1,t_2)=G_{{\bf q}}^{-(0)}(t_1-t_2)+G_{{\bf q}}^{-(1)}(t_1,t_2)\ ,
\end{equation}

\noindent where $G_{{\bf q}}^{-(0)}(t_1-t_2)$ is the solution of the first Kadanoff-Baym equation when only equilibrium propagators are taken into account, and $G_{{\bf q}}^{-(1)}(t_1,t_2)$ is the nonequilibrium propagator when nonequilibrium self-energies are considered (see fig. \ref{feynman1}). Notice that ${\bf q}$ refers to the thermal bath momentum, and ${\bf k}$ to the nonequilibrium field momentum. This type of expansion can only be achieved if the self-energy can be written as 

\begin{equation}
\Sigma_{{\bf q}}^-(t_1,t_2)=\Sigma_{{\bf q}}^{-(0)}(t_1-t_2)+\Sigma_{{\bf q}}^{-(1)}(t_1,t_2)\ .
\end{equation}

\noindent It is easy to see that the prefix 1 is a higher order in a perturbation series in comparison with the prefix 0. Now, introducing the above equations in (\ref{kbeG1}), one obtains
\begin{eqnarray}
\Boxq G^{-(1)}_{\bf{q}}(t_{1},t_{2})=&-&\int_{t_{2}}^{t_{1}} d t' [\Sigma^{-(0)}_{\bf{q}}(t_1-t')G^{-(1)}_{\bf{q}}(t',t_2)\nonumber\\
&&\hspace{1.3cm}+\Sigma^{-(1)}_{\bf{q}}(t_1,t')G^{-(0)}_{\bf{q}}(t'-t_2)]\ ,
\end{eqnarray} 
where we have cancelled out the equilibrium components and we have neglected higher order in the perturbation series. Rearranging the above equation we obtain

\begin{equation}\label{inhomo}
\Boxq G^{-(1)}_{\bf{q}}(t_{1},t_{2})
+\int_{t_{2}}^{t_{1}} d t' \Sigma^{-(0)}_{\bf{q}}(t_1-t') G^{-(1)}_{\bf{q}}(t',t_2)=\xi (t_1,t_2)\ ,
\end{equation} 
with
\begin{equation}
\xi(t_1,t_2)=-\int_{t_{2}}^{t_{1}} d t' \Sigma^{-(1)}_{\bf{q}}(t_1,t') G^{-(0)}_{\bf{q}}(t'-t_2)\ .
\end{equation}
Equation (\ref{inhomo}) resembles the KBE for the statistical propagator in (\ref{kb2}) calculated in the previous section. It has a homogeneous part (left-hand side) and an inhomogeneous part (right-hand side). The solution for the spectral function is given by

\begin{equation}
 G_{{\bf q}}^{-(1)}(t_1,t_2)= G_{H,{\bf q}}^{-}(t_1,t_2)+ G_{I,{\bf q}}^{-}(t_1,t_2)\ .
\end{equation}
Here the subscript $H$ stands for homogeneous solution and $I$ for inhomogeneous solution. We can notice that the homogeneous solution satisfy the same equation as the equilibrium solution, i.e.
\begin{equation}
 G_{H,{\bf q}}^{-}(t_1,t_2)= G_{{\bf q}}^{-(0)}(t_1-t_2)\ ,
\end{equation}
while the inhomogeneous solution takes the form

\begin{equation}
 G^{-}_{I,\bf{q}}(t_{1},t_{2})
=\int_{t_{2}}^{t_{1}} d t'  G^{-(0)}_{\bf{q}}(t_1-t')\xi(t',t_2)\ ,
\end{equation} 

\noindent which can be written in a general form with no restrictions on $t_1$ and $t_2$ 
\begin{eqnarray}\label{fullsol}
 G^{-}_{I,\bf{q}}(t_{1},t_{2})=&&\int_{t_i}^{t_1}dt' \int_{t_i}^{t_2}dt''[ G^{-(0)}_{\bf{q}}(t_1-t')\Sigma^{-(1)}(t',t'') G^{-(0)}_{\bf{q}}(t''-t_2)]\nonumber\\
&-&\int_{t_i}^{t_1}dt' \int_{t_i}^{t'}dt''[ G^{-(0)}_{\bf{q}}(t_1-t')\Sigma^{-(1)}(t',t'') G^{-(0)}_{\bf{q}}(t''-t_2)]\nonumber\\
&-&\int_{t_i}^{t_2}dt'' \int_{t_i}^{t''}dt'[ G^{-(0)}_{\bf{q}}(t_1-t')\Sigma^{-(1)}(t',t'') G^{-(0)}_{\bf{q}}(t''-t_2)]\ .
\end{eqnarray}

This equation, as it is normal with nonequilibrium phenomena, will depend on the initial conditions given by the initial time $t_i$, although it will explicitly disappear. We will see that because of the condition $t_1,t_2>t_i$, the initial conditions will have an effect on the phase-space when a Fourier transformation is performed on the time difference $|t_1-t_2|$. Now, for the case where $t_1>t_2$ the time integration will became 
\begin{equation}
\int_{t_i}^{t_1}dt' \int_{t_i}^{t_2}dt''-\int_{t_i}^{t_1}dt' \int_{t_i}^{t'}dt''-\int_{t_i}^{t_2}dt'' \int_{t_i}^{t''}dt'=-\int_{t_2}^{t_1}dt' \int_{t_2}^{t'}dt''\ ,
\end{equation} 

\noindent giving us a result
\begin{equation}\label{solt1mayt2}
 G^{-}_{I,\bf{q}}(t_{1},t_{2})
=-\int_{t_{2}}^{t_{1}} d t'\int_{t_{2}}^{t'} d t''   G^{-(0)}_{\bf{q}}(t_1-t') \Sigma^{-(1)}_{\bf{q}}(t',t'') G^{-(0)}_{\bf{q}}(t''-t_2)\ .
\end{equation} 

\noindent And for the case $t_2>t_1$ we have

\begin{equation}
\int_{t_i}^{t_1}dt' \int_{t_i}^{t_2}dt''-\int_{t_i}^{t_1}dt' \int_{t_i}^{t'}dt''-\int_{t_i}^{t_2}dt'' \int_{t_i}^{t''}dt'=-\int_{t_1}^{t_2}dt'' \int_{t_1}^{t''}dt'\ ,
\end{equation} 

\noindent giving us a result for this case

\begin{equation}\label{solt2mayt1}
 G^{-}_{I,\bf{q}}(t_{1},t_{2})
=-\int_{t_{1}}^{t_{2}} d t''\int_{t_{1}}^{t''} d t'   G^{-(0)}_{\bf{q}}(t_1-t') \Sigma^{-(1)}_{\bf{q}}(t',t'') G^{-(0)}_{\bf{q}}(t''-t_2)\ ,
\end{equation} 

If the self-energy depends on the difference of times $\Sigma^{-(1)}_{\bf{q}}(t',t'')\rightarrow \Sigma^{-(1)}_{\bf{q}}(t'-t'')$ one will obtain the expected $ G^{-}_{mem,\bf{q}}(t_{1},t_{2})\rightarrow G^{-}_{mem,\bf{q}}(t_{1}-t_{2})$. In order to check this,  one needs to take the Fourier transform on the time difference and then integrate on $t'$ and $t''$.

In order to continue we will need to calculate the self-energy in order to have a full solution to the nonequilibrium spectral function.

\subsection*{Nonequilibrium self-energy}
The self-energy to include in the inhomogeneous solution (\ref{solt2mayt1}) depends explicitly on the nonequilibrium propagators of $\phi$ (the full list of propagators is given in appendix \ref{appA}). One can first see that the propagators can be always written as a sum of an equilibrium term and a nonequilibrium one, this comes from the fact that the equilibrium condition is already included in the propagators
\begin{equation}
\Delta_{\bf{k}}(t_1,t_2)=\Delta_{eq,\bf{k}}(t_1-t_2)+\Delta_{non,\bf{k}}(t_1,t_2)\ .
\end{equation}
\noindent This leads to a separation of the self-energy in an equilibrium and nonequilibrium part, therefore, a separation in the spectral function
\begin{eqnarray}\label{gminusnon}
G^{-}_{I,\bf{q}}(t_{1},t_{2})=&-&\int_{t_2}^{t_1} dt'\int_{t_2}^{t'}dt'' \Big[G^{-(0)}_{\bf{q}}(t_1-t') \Sigma^{-(1)}_{eq,\bf{q}}(t'-t'')G^{-(0)}_{\bf{q}}(t''-t_2)\nonumber\\
&&\hspace{2.7cm}+\;G^{-(0)}_{\bf{q}}(t_1-t') \Sigma^{-(1)}_{non,\bf{q}}(t',t'')G^{-(0)}_{\bf{q}}(t''-t_2)\Big]\ .
\end{eqnarray}
The first line in the right-hand side of the above equation is already included in the homogeneous term because, as we mentioned above, the homogeneous solution for $G^{-}_{\bf{q}}$ with the interaction with $\phi$ is also present in the nonequilibrium propagators.

To proceed, we will use the definition of the antisymmetric self-energy
\begin{equation}\label{selfminus-}
\Sigma_{non,\bf{q}}^{-}(t',t'')=i(\Sigma_{non,{\bf q}}^{>}(t',t'')-\Sigma_{non,{\bf q}}^{<}(t',t''))\ .
\end{equation}
\noindent Both $\gtrless$ 1-loop self-energies can be calculated with the appropriate propagators (see fig. \ref{feynman1})
\begin{equation}
\Sigma_{non,{\bf q}}^{\gtrless}(t',t'')=g^2 \int\frac{d^3k}{(2\pi)^3}[\Delta^{\gtrless}_{non,\bf{k}}(t',t'')G^{\gtrless}_{\bf{q'}}(t',t'')]\ ,
\end{equation}

\noindent Notice that here ${\bf q'}={\bf k}-{\bf q}$ comes from momentum conservation in each vertex. Inserting both equations in (\ref{selfminus-}) we get
\begin{equation}
\Sigma_{non,{\bf q}}^{-}(t',t'')=ig^2 \int\frac{d^3k}{(2\pi)^3}[\Delta^{>}_{non,\bf{k}}(t',t'')G^{>}_{\bf{q'}}(t',t'')-\Delta^{<}_{non,\bf{k}}(t',t'')G^{<}_{\bf{q'}}(t',t'')]\ .
\end{equation}
\noindent Using the property $\Delta^{>}_{non,\bf{k}}(t',t'')=(\Delta^{<}_{non,\bf{k}}(t',t''))^*$, we see that

\begin{equation}
\Sigma_{non,{\bf q}}^{-}(t',t'')=-2g^2 \int\frac{d^3k}{(2\pi)^3}{\rm Im}[\Delta^{>}_{non,\bf{k}}(t',t'')G^{>}_{\bf{q'}}(t',t'')]\ ,
\end{equation}

\noindent which can be written as
\begin{eqnarray}\label{sigmasss}
\Sigma_{non,{\bf q}}^{-}(t',t'')=&&-2g^2  \int\frac{d^3k}{(2\pi)^3}\Big\{ {\rm Re}[\Delta^{>}_{non,\bf{k}}(t',t'')]{\rm Im}[G^{>}_{\bf{q'}}(t',t'')]\nonumber\\
&&\hspace{3cm}+{\rm Im}[\Delta^{>}_{non,\bf{k}}(t',t'')]{\rm Re}[G^{>}_{\bf{q'}}(t',t'')]\Big\}\ .
\end{eqnarray}

The Introduction of the propagators that we obtained for the $\phi$ field (appendix \ref{appA}) needs to be performed separately for the case $t_1>t_2$ and $t_1<t_2$. In the following we will show the former case. The computation for the second case is straightforward. After some simple algebra we obtain, by using equation (\ref{sigmasss}) in (\ref{gminusnon}), the inhomogeneous solution of $G^{-(1)}_{\bf q}$
\begin{eqnarray}
G^{-(1)}_{I,\bf{q}}(t_1,t_2) &=& 
 g^2\int_{t_2}^{t_1} dt'\int_{t_2}^{t'}dt^{''}
 \int \frac{d^3k}{(2\pi)^3}
\frac{1}{\omega_q^2 \; \omega_k \; \omega_{q'}} 
\nonumber \\
& \times & \sin \left(\omega_q (t_1-t') \right) \sin \left(\omega_q (t''-t_2) \right)
{\rm e}^{-\frac{\Gamma_q}{2}\mid t_1-t'\mid}
{\rm e}^{-\frac{\Gamma_q}{2}\mid t''-t_2\mid}
{\rm e}^{-\frac{\Gamma_{q'}}{2}\mid t'-t'' \mid}
\nonumber \\
&\times & \bigg[ \frac{\coth}{2}\left(\frac{\beta \omega_k}{2} \right)
 \sin\left(\omega_{q'} (t'-t'') \right)
 \cos \left(\omega_k (t'-t'') \right) {\rm e}^{-\frac{\Gamma_{k}}{2} (t'-t'')}
 \nonumber \\
& & + \frac{\coth}{2}\left(\frac{\beta \omega_{q'}}{2} \right)
 \cos\left(\omega_{q'} (t'-t'') \right)
 \sin \left(\omega_k (t'-t'') \right) {\rm e}^{-\frac{\Gamma_{k}}{2} (t'-t'')}
 \nonumber \\
&& - f^B(\omega_k)
\cos\left(\omega_k (t'-t'') \right) \sin\left(\omega_{q'} (t'-t'') \right) {\rm e}^{-\frac{\Gamma_{k}}{2} (t'+t'')} 
\bigg]
\ ,
\end{eqnarray}

\noindent where $\omega_{k}=\sqrt{{\bf k}^2+M^2}$, $\omega_{q}=\sqrt{{\bf q}^2+m_{\chi}^2}$ and $\omega_{q'}=\sqrt{{\bf q'}^2+m_{\chi}^2}$. Notice that both times $t_1,t_2\gtrsim 0$, this can be interpreted as an initial condition of the spectral function, this feature will be important when one wants to define an energy of the nonequilibrium field. 

Although the integration in the time coordinates appears to be simple, one needs to be careful with the absolute values in the exponentials. For the case $t_1>t_2$ it is easy to check that $t'>t''$.

The integration over time will result on a pole structure of a combination of $\omega_q$, $\omega_q'$, $\omega_k$ and the decay widths $\Gamma_q$, $\Gamma_q'$ and $\Gamma_k$. In the following we will consider the case where $\Gamma_q=\Gamma_q'$, as both are the same type of particle of the thermal bath $\chi$ and no flavour effects are considered. The integration can be performed by changing the sines and cosines to exponential, which will give a combination of the form

\begin{equation}
\mathcal{J}=e^{\left(i(\alpha\omega_{k}+\beta\omega_{q}+\gamma\omega_{q'})+\frac{\Gamma_{k}}{2}\right)\tilde{t}}\ ,
\end{equation}

\noindent where $\tilde{t}=t',t''$ and $\alpha,\beta,\gamma=\pm 1$.

As the same in previous sections, we will focus on the scenario where the mass of the field $\phi$ is much larger than the mass of the particles in the bath $\chi$, i.e. $M\gg m_{\chi}$. As a result the pole structure of the function $\mathcal{J}$ after integrating $\tilde{t}$ that will have a bigger contribution will be given by $\mathcal{I}$

\begin{equation}
\mathcal{I}=(\omega_{k}-\omega_{q}-\omega_{q'})^2+\frac{\Gamma_k^2}{4}\ ,
\end{equation}

\noindent where for massles $\chi$ fields one has $\omega_{q}=|{q}|$, $\omega_{q'}=|{\bf q'}|=|{\bf k}-{\bf q}|$ and $\omega_{k}=\sqrt{{\bf k}^2+m^2}$. For small or large ${\bf k}$ one can easily see that $\mathcal{I}\sim \Gamma_{k}$ so that the spectral function will be of order $\sim 1/\Gamma_{k}$.
After performing the time integration and maintaining the $1/\Gamma_{k}$ terms one obtains
\begin{eqnarray}\label{gnon0}
G^{-(1)}_{non,{\bf q}}(t_1,t_2) &=& \frac{g^2}{8}
\int \frac{d^3k}{(2\pi)^3} \frac{1}{\omega_q^2 \; \omega_k \; \omega_{q'}} f^B(\omega_k)
\left[ \frac{1}{\omega^2_p+\frac{\Gamma_k^2}{4}} \right]
{\rm e}^{-\frac{\Gamma_q}{2}(t_1-t_2)} \nonumber \\
&\times &\left[ \frac{\omega_p}{\Gamma_k}\cos[(t_1-t_2)\omega_q]
\left({\rm e}^{-\Gamma_k t_2}-{\rm e}^{-\Gamma_k t_1}\right)\right. \nonumber\\
& &+\frac{1}{2} \sin[(t_1-t_2)\omega_q]
\left({\rm e}^{-\Gamma_k t_2}+{\rm e}^{-\Gamma_k t_1}\right)  \nonumber \\
& & \left.-\sin[(\omega_p+\omega_q)(t_1-t_2)]{\rm e}^{-\frac{\Gamma_k}{2} (t_1+t_2)} \right]\ ,
\end{eqnarray}

\noindent with $\omega_{p}=\omega_{k}-\omega_{q}-\omega_{q'}$. Another way of writing the above equation is to perform the time change into the time difference $y=t_1-t_2$ and the center of mass time $t=(t_1+t_2)/2$ which gives

\begin{eqnarray}\label{gnon1}
G^{-(1)}_{non,{\bf q}}(y,t) &=& \frac{g^2}{8}
\int \frac{d^3k}{(2\pi)^3} \frac{1}{\omega_q^2 \; \omega_k \; \omega_{q'}} f^B(\omega_k)
\left[ \frac{1}{\omega^2_p+\frac{\Gamma_k^2}{4}} \right]
{\rm e}^{-\frac{\Gamma_q}{2}y}{\rm e}^{-\frac{\Gamma_k}{2}t} \nonumber \\
&\times &\left[ \frac{2\omega_p}{\Gamma_k}\cos[\omega_q y]\sinh \left(\frac{\Gamma_k y}{2}\right)
+\sin[\omega_q y]
\cosh \left(\frac{\Gamma_k y}{2}\right)-\sin[(\omega_p+\omega_q)y]\right]\ .\nonumber\\
\end{eqnarray}
	
\section{Sum rule}
One of the most important aspects of the spectral function that arises from the boundary conditions, called the sum rule, is the fact that the area of the spectral function in the frequency phase-space ($\omega$) is constant. Meaning that, for example, any change that modifies the height of the spectral function will lead automatically to a broadness of its width and vice-versa. We will show how the nonequilibrium corrections does not affect such sum rule, and it can modify the spectral function without changing its area.
Using the border conditions shown in eqs. (\ref{con1})-(\ref{con3}) one can write for any spectral function $\sigma$
\begin{eqnarray}
\sigma(t_1,t_2)|_{t_1=t_2}&=&0\ ,\\
\partial_{t_1}\sigma_{\bf }(t_1,t_2)|_{t_1=t_2}=-\partial_{t_2}\sigma(t_1,t_2)|_{t_1=t_2}&=&1\ ,\\
\partial_{t_1}\partial_{t_2}\sigma(t_1,t_2)|_{t_1=t_2}&=&0\ .
\end{eqnarray}

\noindent We can split the spectral function into an equilibrium and nonequilibrium contribution 

\begin{equation}
\sigma(t_1,t_2)=\sigma_{eq}(t_1-t_2)+\sigma_{non}(t_1,t_2)\ .
\end{equation}

Notice the dependence on the difference of time for the equilibrium spectral function. One can easily verify that the border conditions is equally satisfied for the nonequilibrium part. This is done by taking the general solution (\ref{fullsol}) or solutions (\ref{solt1mayt2}) and (\ref{solt2mayt1}). The first condition is justified as the spectral function is antisymmetric. The second condition can be satisfy by using our solution for the nonequilibrium contribution in eq. (\ref{fullsol}), or the exact solution for this model given by eq. (\ref{gnon1}).

In order to visualize the physical properties of the border conditions we need to perform a Fourier transform in the variable $y$. Although one can naively think as this transform as going to energy phase-space, we should emphasize that this quantity is not conserved under the integral in nonequilibrium scenarios, that means that it should not be treated as the energy of the thermal bath. Now, one can write

\begin{equation}
\sigma(y,t)=i\int_{\mathcal{F}}\frac{d\omega}{2\pi} \rho(\omega,t)e^{-i\omega y}\ ,
\end{equation}

\noindent where $\int_{\mathcal{F}}$ denotes the appropriate domain of integration and $\rho(\omega,t)=-i\sigma(\omega,t)$ is defined to be a real quantity. Notice that the dependence on $\omega$ means the Fourier transform of the $y$ variable. The first boundary condition implies that at $y=0$ we have

\begin{equation}
\int_{\mathcal{F}}d\omega \rho(\omega,t)=0\ ,
\end{equation}

\noindent which shows the antisymmetric property of the spectral function $\rho(\omega,t)=-\rho(-\omega,t)$. The second condition must be applied more carefully. For instance, the time $t$ dependance for an equilibrium spectral function will disappear $\rho_{eq}(\omega,t)\rightarrow\rho_{eq}(\omega)$\footnote{This is the same as taking the time $t\rightarrow\infty$, as the spectral function should thermalize when times goes to infinity.} which gives

\begin{equation}\label{eqsumrule}
\int_{\mathcal{F}}d\omega\; \omega\rho_{eq}(\omega)=1\ .
\end{equation}

\noindent This last equation is usually known as the sum rule of the equilibrium spectral function. It relates the peak of the spectral function to the decay width associated to the field. When the field is out-of-equilibrium one needs to be more careful with the time derivative, this gives

\begin{equation}
\partial_{t_1}\sigma(y,t)|_{y=0}=\int_{\mathcal{F}}\frac{d\omega}{2\pi} e^{-i\omega y}\omega\rho(\omega,t)|_{y=0}+i\int_{\mathcal{F}}\frac{d\omega}{2\pi} e^{-i\omega y}\partial_{t_1}\rho(\omega,t)|_{y=0}=1\ .
\end{equation}

\noindent The dependance on $t$ makes the new term appears. Because of the nature of the Kadanoff-Baym equation, one can write the $\omega$-dependent  spectral function as an equilibrium plus a nonequilibrium part

\begin{equation}
\rho(\omega,t)=\rho_{eq}(\omega)+\rho_{non}(\omega,t)\ ,
\end{equation}

\noindent for $y=0$ it will follow that 

\begin{equation}
\int_{\mathcal{F}}\frac{d\omega}{2\pi}\; \omega\rho_{eq}(\omega)+\int_{\mathcal{F}}\frac{d\omega}{2\pi}\; \omega\rho_{non}(\omega,t)+i\int_{\mathcal{F}}\frac{d\omega}{2\pi}\; \partial_{t_1}\rho_{non}(\omega,t)=1\ .
\end{equation}

\noindent The equilibrium spectral function satisfies (\ref{eqsumrule}), the above equation can be written as

\begin{equation}
\int_{\mathcal{F}}\frac{d\omega}{2\pi}\; \omega\rho_{non}(\omega,t)=-i\int_{\mathcal{F}}\frac{d\omega}{2\pi}\; \partial_{t_1}\rho_{non}(\omega,t)\ .
\end{equation}

\noindent The right hand side is obviously zero taking into account the antisymmetric nature of the $\rho$ function in $\omega$. Leaving us the sum rule for the nonequilibrium part of the spectral function 

\begin{equation}
\int_{\mathcal{F}}\frac{d\omega}{2\pi}\; \omega\rho_{non}(\omega,t)=0\ .
\end{equation}

Although the last equation appears to violate the antisymmetric property of the spectral function, one needs to notice that this equality is true if and only if it vanishes in the positive branch of $\omega$ as well as in the negative part, i.e.

\begin{equation}\label{sumrulenon}
\int_{0}^{\omega_+}\frac{d\omega}{2\pi}\; \omega\rho_{non}(\omega,t)=\int_{\omega_-}^{0}\frac{d\omega}{2\pi}\; \omega\rho_{non}(\omega,t)=0\ ,
\end{equation}

\noindent where $\omega_\pm$ is the corresponding upper and lower limit in the Fourier domain of the frequency $\omega$. The new sum rule of the full nonequilibrium spectral function can be written as 

\begin{equation}\label{neqsumrule}
\int_{\mathcal{F}}d\omega\; \omega\rho(\omega,t)=1\ ,
\end{equation}

\noindent with the nonequilibrium part satisfying eq. (\ref{sumrulenon}).

\section{Disscusion}

Equations (\ref{gnon0}) and (\ref{gnon1}) show the oscillatory behavior of the nonequilibrium spectral function with exponential suppression on the $y$ and $t$ time coordinates. The latter is an expected behavior for the nonequilibrium part as it should thermalize as time goes to infinity. Notice that the temperature dependance appears as usual on the Bose-Einstein distribution function $f^B(\omega_{k})$ The thermal effect are relevant when $T\gtrsim M/\sqrt{2}$. This means that nonequilibrium effects will only be relevant if the temperature is of the order of the scale of the system, which in this case the mass of the nonequilibrium field.

\begin{figure}[htbp] 
   \centering
   \includegraphics[width=4in]{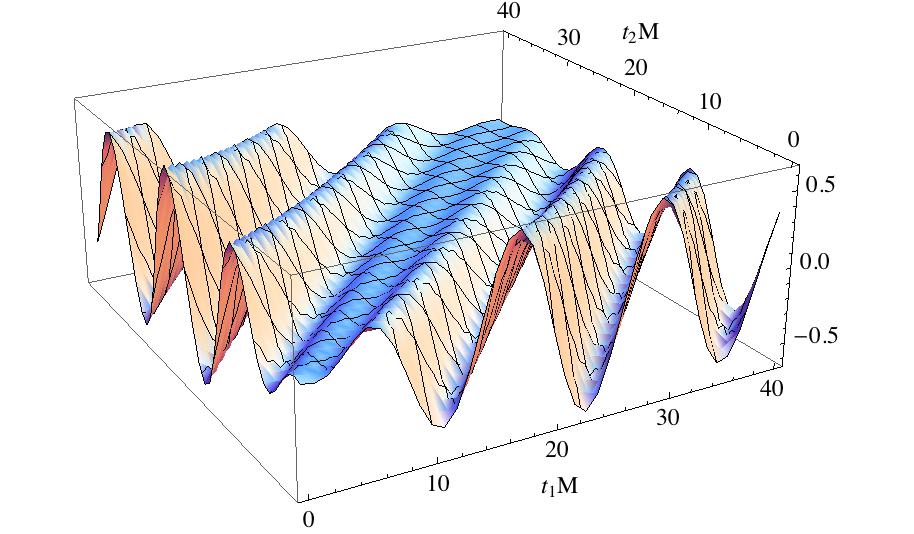} 
   \caption{Nonequilibrium contribution to the spectral function $G^-_{non,{\bf q}}(t_1,t_2)$ with $T=3M$ and $\omega_{\bf q}=0.5M$}
   \label{1loopnon}
\end{figure}

Remember that the full solution for the spectral function is the sum of the nonequilibrium solution with the equilibrium one, so effects will not only depend on the temperature but also on the square of the coupling. The fact that $\Gamma_{\bf k}$ as shown in eq. (\ref{gammadef}) also depends on the square of the coupling implies that nonequilibrium effects can be relevant since they go as $ g^2/\Gamma_{\bf k}$.
Equations (\ref{gnon0}) and (\ref{gnon1}) satisfies the border conditions in (\ref{con1})-(\ref{con3}) and the sum rule in (\ref{sumrulenon}), which means that the nonequilibrium effects can modify the spectral function height but maintaining its area intact. Moreover, these effects can have big repercussion if a second field is involved in the system.\\


The momentum integration must be performed numerically as shown in appendix \ref{appB}. The three dimensional momentum integration can be carefully calculated with the projection of the momenta ${\bf q}$ and ${\bf k}$ through a mixing angle $\theta$. Although the decay widths depend on the momentum of each field, we will for simplicity consider them as constant. This can be justified by looking at the integration window. For example, from the definition (\ref{gammadef}) one can see that $\Gamma_{k}$ is supported near the pole of the propagator, that is near the quasi-particle effective mass and can in principle be taken as constant. The result of the integration can be seen in fig. \ref{1loopnon} with the time variables $t_1$ and $t_2$. From fig. \ref{1loopnon} one can observe the asymmetrical behavior of the spectral function, as well as the oscillatory behavior and exponential suppression in both axes.

\begin{figure}[h!] 
   \centering
   \includegraphics[width=5in]{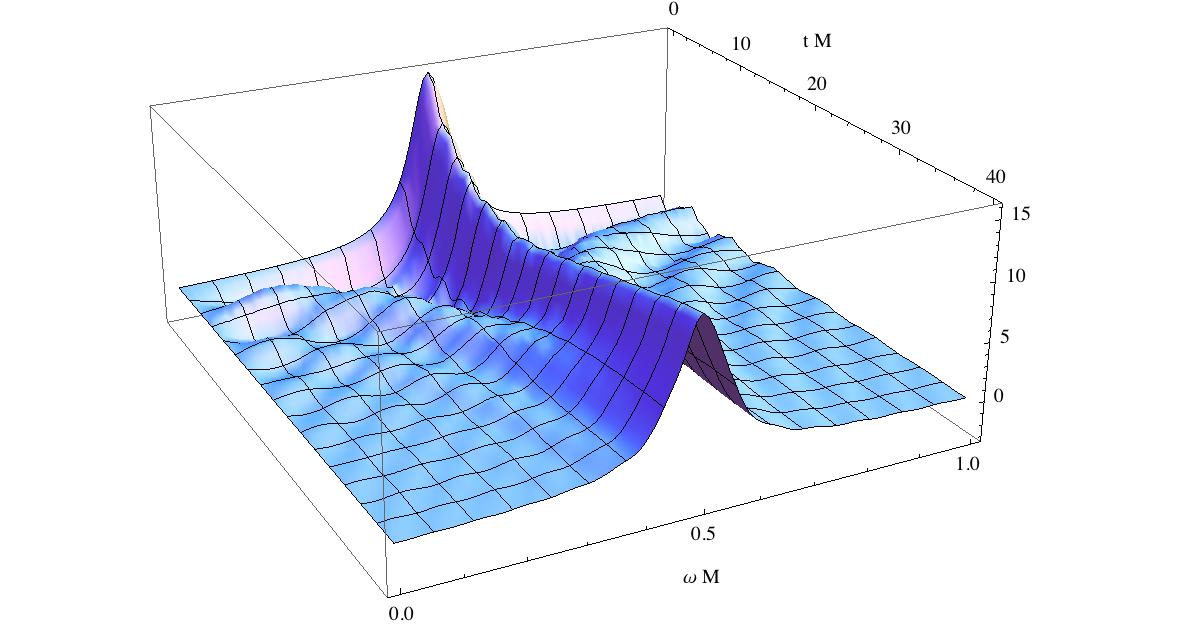} 
   \caption{Full spectral function $G^-_{\bf q}(\omega,t)=G^-_{eq,\bf{q}}(\omega)+G^-_{non,{\bf q}}(\omega,t)$}
   \label{fullrho}
\end{figure}

Although, as we mentioned before, the initials conditions do not appear explicitly in the solutions, one needs to have in mind that these solutions are only valid in the range $|y|\leq 2t$, restriction that comes from the fact that we imposed $t_1,t_2\geq 0$.

A natural way to physically understand the properties of the spectral function, is to go to frequency space through a Fourier transformation of the $y=t_1-t_2$ coordinate. In equilibrium, this transformation corresponds to shift to the energy phase space of a pseudo-particle with its corresponding thermal corrections to the mass. In nonequilibrium scenarios it is still possible to define a frequency $\omega$ related to $y$. However, this new variable $\omega$ is not a conserved quantity such as energy in equilibrium. Despite the above, we can still use this new parameter to describe the thermalization.

Introducing the variable $\omega$ as the Fourier transform of $y$ will give the full spectral function in fig. \ref{fullrho}, that is the sum of the equilibrium solution to the nonequilibrium one. This figure gives a very good description of the thermalization and back-reaction suffered by the thermal bath. It starts as a normal Breit-Wigner at $t=0$ and goes out-of-equilibrium as the $\phi$ field is been created. If the coupling or temperature is big enough, the effect can be substantial. The wiggles observed in the figure is the way the nonequilibrium effects satisfy the sum rule discussed above. \\

\begin{figure}[h!] 
   \centering
   \includegraphics[width=5in]{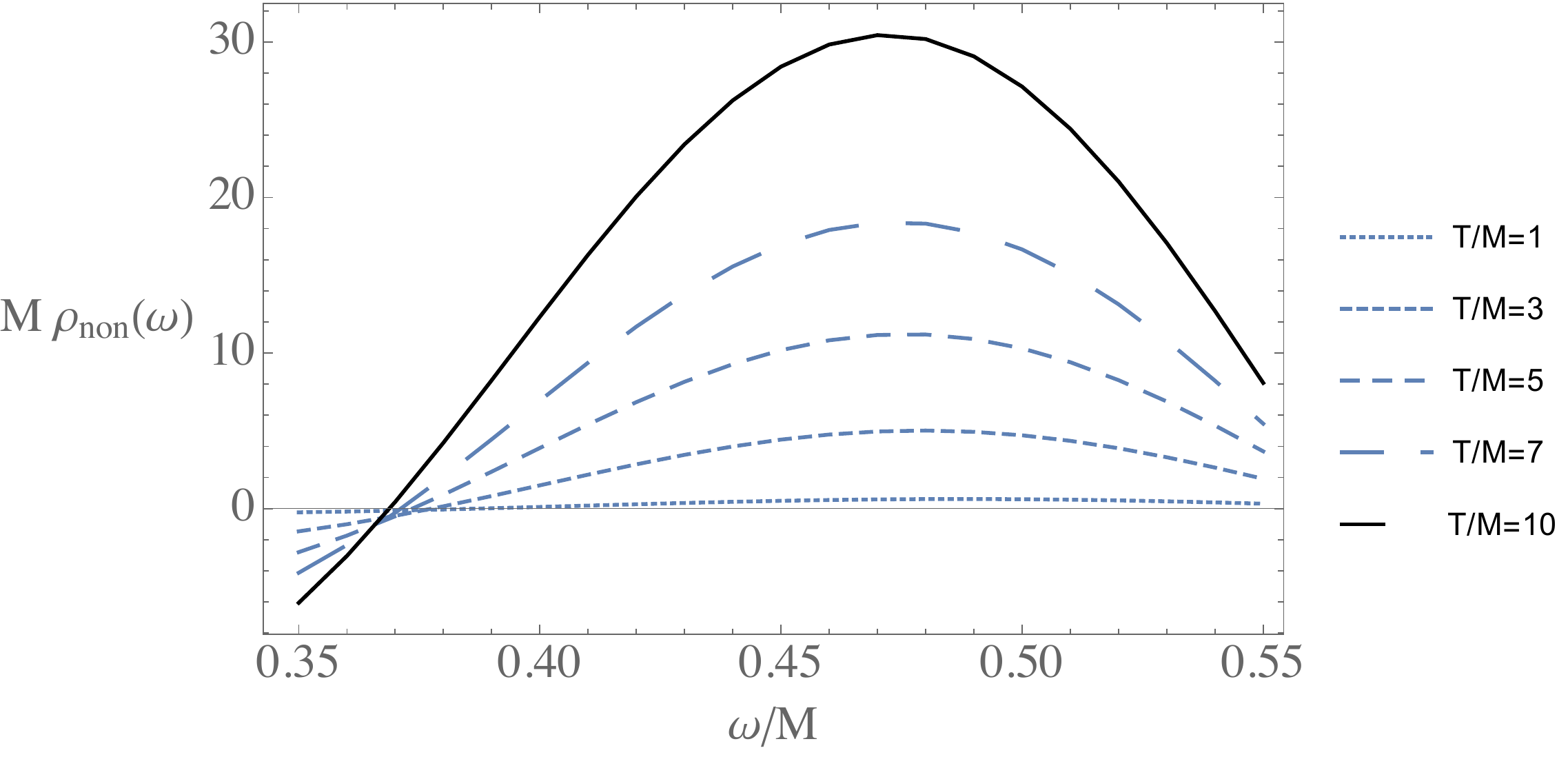} 
   \caption{Nonequilibrium spectral function $\rho^{non}_{{\bf q}}(\omega)$ in terms of the bath frequency $\omega$ for different values of the temperature. Time $t=12/M$ and $q=0.5M$}
   \label{fouriervsw}
\end{figure}

Fig. \ref{fouriervsw} shows a transversal cut of the nonequilibrium contribution for a fixed time ($t=12/M$) for different temperatures. As the temperature raises, the effect of the backreaction increases as well. However, when the temperature is high enough, effects as thermal masses or corrections to the coupling and width must be taken into account. One particular point of interest here is the shift of the pole when temperature increases. This pole shift correction is similar as the one that appears with the inclusion of thermal masses, it indicates that the efficient energy where the process takes place is not at the mass equilibrium pole but it is shifted due to the quantum corrections that need to satisfy the sum rule.\\ 

\begin{figure}[htbp] 
   \centering
   \includegraphics[width=5in]{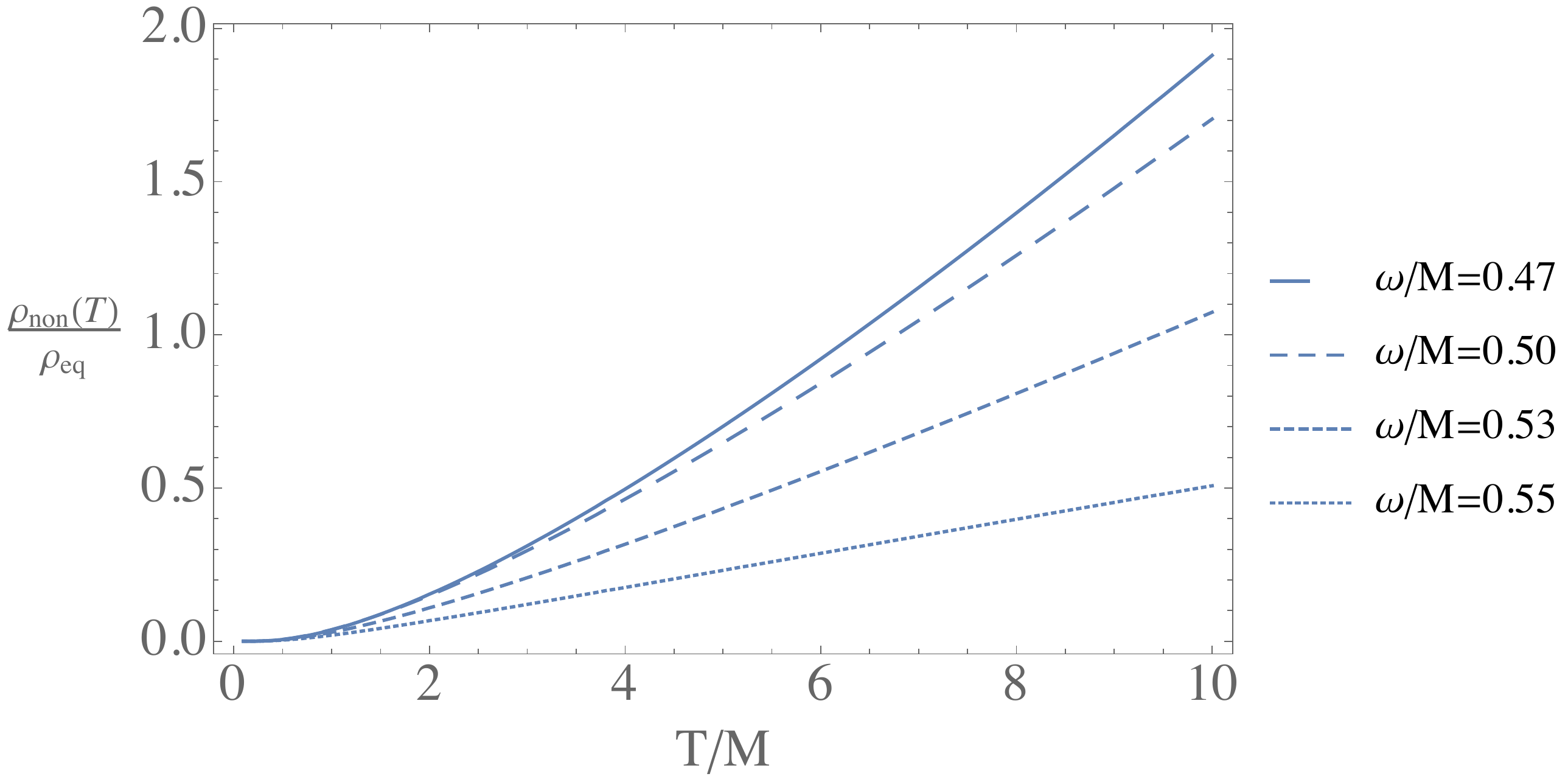} 
   \captionof{figure}{Nonequilibrium spectral function $\rho^{non}_{{\bf q}}(T)/\rho^{eq}_{{\bf q}}$ in terms of the temperature for a different values of $\omega$, for fixed time $t=12/M$ and $q=0.5M$.}
   \label{fouriervsT}
\end{figure}

To show how big this new nonequilibrium corrections are with respect to the equilibrium spectral density at its maximum value, we can focus on different values $\omega$ as seen in fig. \ref{fouriervsT}. At $\omega=0.47M$ it exhibits the biggest increment as the temperature increases, confirming a direct consequence of the pole moving from equilibrium. This shift starts to occur occurs for $T\gtrsim M$. For values of $T\gtrsim 4M$, the correction is significant compared to the equilibrium spectral density and increases linearly with $T$.

\subsection*{Aknowledgment}	
JCR and SM would like to thank FONDECYT 1150471. SM would like to thank FONDECYT  11130118. JCR would like to thank FONDECYT 1130056 and FONDECYT 1150847. We would also like to thank Marco Drewes for insightful discussions.
	
\appendix
\section*{Appendices}
\numberwithin{equation}{section}
\section{Propagators}\label{appA}
Equilibrium propagators for the $\phi$ field

\begin{eqnarray}
\Delta^{(0),-}_{ \textbf{k}}(y)&=&\frac{1}{\omega_{ \textbf{k}}}\sin(\omega_{
\textbf{k}}y)e^{-\frac{\Gamma_{\bf k}}{2}|y|}\;,\nonumber\\
\Delta^{(0),+}_{ \textbf{k}}(y)&=&\frac{1}{2\omega_{ \textbf{k}}}\cos(\omega_{
\textbf{k}}y)\coth\left(\frac{\beta\omega_{ \textbf{k}}}{2}\right)e^{-\frac{\Gamma_{\bf k}}{2}|y|}\;,\nonumber\\
\Delta^{(0),11}_{ \textbf{k}}(y)&=&\frac{1}{2\omega_{ \textbf{k}}}\left(\coth\left(\frac{\beta\omega_{
\textbf{k}}}{2}\right)\cos(\omega_{ \textbf{k}}y)-i\sin(\omega_{ \textbf{k}}|y|)\right)e^{-\frac{\Gamma_{\bf k}}{2}|y|}\;,\nonumber\\
\Delta^{(0),22}_{ \textbf{k}}(y)&=&(\Delta^{(0),11}_{ \textbf{k}}(y))^*\;,\nonumber\\
\Delta^{(0),>}_{ \textbf{k}}(y)&=&\frac{1}{2\omega_{
\textbf{k}}}\left(\cos(\omega_{ \textbf{k}}y)\coth\left(\frac{\beta\omega_{
\textbf{k}}}{2}\right)-i\sin(\omega_{ \textbf{k}}y)\right)e^{-\frac{\Gamma_{\bf k}}{2}|y|}\;,\nonumber\\
\Delta^{(0),<}_{ \textbf{k}}(y)&=&(\Delta^{(0),>}_{ \textbf{k}}(y))^*\;.
\end{eqnarray}

\noindent Nonequilibrium propagators for the $\phi$ field

\begin{eqnarray}
\Delta^{(1),-}_{ \textbf{k}}(y)&=&\frac{1}{\omega_{ \textbf{k}}}\sin(\omega_{
\textbf{k}}y)e^{-\frac{\Gamma_{\bf k}}{2}|y|}\;,\nonumber\\
\Delta^{(1),+}_{ \textbf{k}}(y)&=&\frac{1}{\omega_{ \textbf{k}}}\cos(\omega_{
\textbf{k}}y)\left[\frac{1}{2}\coth\left(\frac{\beta\omega_{ \textbf{k}}}{2}\right)e^{-\frac{\Gamma_{\bf k}}{2}|y|}-f^B(\omega_{ \textbf{k}})e^{-\Gamma_{ \textbf{k}}t}\right]\;,\nonumber\\
\Delta^{(1),11}_{ \textbf{k}}(y)&=&\frac{1}{\omega_{
\textbf{k}}}\cos(\omega_{ \textbf{k}}y)\left(\frac{1}{2}\coth\left(\frac{\beta\omega_{
\textbf{k}}}{2}\right)e^{-\frac{\Gamma_{\bf k}}{2}|y|}-f^B(\omega_{ \textbf{k}})e^{-\Gamma_{ \textbf{k}}t}\right)-\frac{i}{2\omega_{ \textbf{k}}}\sin(\omega_{ \textbf{k}}|y|)e^{-\frac{\Gamma_{\bf k}}{2}|y|}\;,\nonumber\\
\Delta^{(1),22}_{ \textbf{k}}(y)&=&(\Delta^{eq,11}_{ \textbf{k}}(y))^*\;,\nonumber\\
\Delta^{(1),>}_{ \textbf{k}}(y)&=&\frac{1}{\omega_{
\textbf{k}}}\cos(\omega_{ \textbf{k}}y)\left(\frac{1}{2}\coth\left(\frac{\beta\omega_{
\textbf{k}}}{2}\right)e^{-\frac{\Gamma_{\bf k}}{2}|y|}-f^B(\omega_{ \textbf{k}})e^{-\Gamma_{ \textbf{k}}t}\right)-\frac{i}{2\omega_{ \textbf{k}}}\sin(\omega_{ \textbf{k}}y)e^{-\frac{\Gamma_{\bf k}}{2}|y|}\;,\nonumber\\
\Delta^{(1),<}_{ \textbf{k}}(y)&=&(\Delta^{eq,>}_{ \textbf{k}}(y))^*\;.
\end{eqnarray}

\noindent where 

\begin{equation}
f^B(\omega_{ \textbf{k}})=\frac{1}{e^{\beta\omega_{ \textbf{k}}}-1}\;,\nonumber
\end{equation}
and $\beta$ is the inverse of the temperature.

\section{Numerical computation}\label{appB}
Lets rewrite equation (\ref{gnon1})
\begin{equation}
G^{-(1)}_{non,{\bf q}} (y,t)=\frac{g^2}{8}
\int \frac{d^3k}{(2\pi)^3} \frac{1}{\omega_q^2 \; \omega_k \; \omega_{q'}} f^B(\omega_k)
\left[ \frac{1}{\frac{\Gamma_k^2}{4}+\omega^2_p} \right]
{\rm e}^{-\frac{\Gamma_q}{2}|y|}{\rm e}^{-\frac{\Gamma_k}{2}t} F(\omega_p).
\end{equation}
\noindent with $\omega_p=\omega_k-\omega_q-\omega_{q'}$ and the function $F(\omega_p)$ given by
\begin{eqnarray}
F(\omega_p)=\frac{2\omega_p}{\Gamma_k}\cos[\omega_q y]\sinh (\Gamma_k y)
+\sin[\omega_q y]
\cosh (\Gamma_k y)-\sin[(\omega_p+\omega_q)y],
\end{eqnarray}
The momemtum ${\bf q'}={\bf k}-{\bf q}$ is just the difference between the momentum of the eternal bath momentum and the nonequilibrium field $\phi$. So that the integration depends also on the angle between these two momenta. Making the change of variables for a massless bath particle (i.e. $q'=\omega_{q'}$ and $q=\omega_{q}$)
\begin{equation}
q'=|{\bf k}-{\bf q}|=({\bf k}^2+{\bf q}^2-2{\bf k}\cdot{\bf q}\cos\theta)^{1/2},
\end{equation}
\noindent one obtains the following differential for $q'$
\begin{equation}
dq'=-\frac{{\bf k}\cdot{\bf q}}{q'}d(\cos\theta).
\end{equation}
The minimun and maximun values for $q'$ are given by $q'_{\pm}$, where
\begin{equation}
q'_{\pm}=|{\bf k}\pm{\bf q}|.
\end{equation}
Which will be use in the integration
\begin{equation}
\int d^3k=\int k^2dk\int d(\cos\theta)\int d\phi
\end{equation}
The limits for the $k$ integration can be obtained from $\omega_p=\omega_k-q-q'$ with the limits for $q'$, then one obtains $k^2+M^2<(k+2q)^2$, giving finally
\begin{equation}
k>\frac{M^2-4q^2}{4q}.
\end{equation}
Introducing the change of variables in the original integration gives
\begin{eqnarray}
G^{-(1)}_{non,{\bf q}}(y,t) &=&-\frac{g^2}{8}
\int_0^{2\pi} d\phi\int_{\frac{M^2-4q^2}{4q}}^\infty \frac{dk}{(2\pi)^3} \int_{|k-q|}^{|k+q|}dq'\frac{k}{q^3\sqrt{k^2+M^2}} f^B(\sqrt{k^2+M^2})\nonumber\\
&\times&\left[ \frac{1}{\frac{\Gamma_k^2}{4}+(\sqrt{k^2+M^2}-q-q')^2} \right]
{\rm e}^{-\frac{\Gamma_q}{2}|y|}{\rm e}^{-\frac{\Gamma_k}{2}t} F(\sqrt{k^2+M^2}-q-q').\nonumber\\
\end{eqnarray}

\end{document}